\documentclass[twocolumn,prl,showpacs,preprintnumbers,amsmath,amssymb]{revtex4}
\usepackage{color}
\usepackage{bm, graphicx, amsmath}
\usepackage[mathscr]{eucal}

\begin{document}

\title{Constructing a qubit POVM from quantum data}
\author{Mark Hillery$^{1,2}$}
\affiliation{$^{1}$Department of Physics and Astronomy, Hunter College of the City University of New York, 695 Park Avenue, New York, NY 10065 USA \\ 
$^{2}$Physics Program, Graduate Center of the City University of New York, 365 Fifth Avenue, New York, NY 10016}

\begin{abstract}
Given an ensemble of qubits, which we are told consists of a mixture of two pure states, one with probability $\eta_{0}$ and one with probability $\eta_{1}$,  we want to find a POVM that will discriminate between the two states by measuring the qubits.  We do not know the states, and for any given qubit, we do not know which of the two states it is in.  This can be viewed as learning a POVM from quantum data.  Once found, the POVM can be used to separate the remaining qubits in the ensemble into two groups, corresponding to the two states present in the ensemble.  In order to find the POVM, we need more information about the possible states.  We examine several cases.  First, we suppose that we know that the Bloch vectors of the states lie in the $x-z$ plane and their \emph{a priori} probabilities are equal.  We next keep the restriction to the $x-z$ plane, but allow the \emph{a priori} probabilities to be different.  Finally, we consider the case in which the Bloch vectors of the states  have the same $z$ component.
\end{abstract}

\maketitle

\section{Introduction}
Quantum learning is related to classical machine learning, but has many unique aspects.  One can learn a number of different quantum objects, unitary operators \cite{bisio} and measurements \cite{sedlak,guta,sentis,fanizza}, for example.  In most cases there is a training set.  In the case of a unitary operator, one is allowed a certain number of uses of the operator, and in the case of a measurement, one is given examples of the states one wants the measurement to distinguish.  What we want to do here is to see what can be done in the case of learning a measurement in which there is no training set.  This is analogous to unsupervised machine learning.

In unsupervised machine learning, one has data that one would like to arrange into clusters.  Quantum algorithms have been applied to obtain speedups of the unsupervised learning of classical data \cite{Brassard}-\cite{kerenidas} (for reviews of quantum machine learning see \cite{schuld}).  In these works the classical data is converted into quantum states, which can then be processed by a quantum computer.  What, however, can be done if one is presented with data in the form of unknown quantum states?  The first treatment of this kind of quantum unsupervised learning was given in \cite{bagan}.  There one was given a sequence of $N$ particles, each of which is in one of two unknown states, $|\psi_{0}\rangle$ and $|\psi_{1}\rangle$, and one wants to determine the sequence.  The output of this procedure is classical, a sequence of $0$'s and $1$'s, corresponding to the labels of the states, of length $N$ that is the best guess for the sequence of states.

The problem we wish to consider here is related but different, and is the following.  We are given a sequence, or ensemble, of qubits, and we are told that each qubit is in a state $|\psi_{0}\rangle$ with probability $\eta_{0}$ or in a state $|\psi_{1}\rangle$ with probability $\eta_{1}$.  We do not know what $|\psi_{0}\rangle$ or $|\psi_{1}\rangle$ are, and we do not know which state any given qubit is in.  Our task is to construct a POVM that will discriminate between $|\psi_{0}\rangle$ and $|\psi_{1}\rangle$ by performing measurements on the ensemble.  This POVM will make errors, but it can be chosen to minimize those errors.   From the theory of state discrimination, we know that the POVM that optimally discriminates these states with minimum error is a projective measurement \cite{helstrom,bergou}.  The POVM, once constructed, can be used to divide the remaining data into two groups, each group corresponding to one of the states in the ensemble.

Additional information beyond that specified above is necessary to accomplish our goal for the following reason.  The density matrix describing this ensemble of qubits is 
\begin{equation}
\label{density}
\rho = \eta_{0} |\psi_{0}\rangle\langle \psi_{0}| + \eta_{1} |\psi_{1}\rangle\langle \psi_{1}| .
\end{equation} 
This density matrix can be determined by measurements made on the ensemble.  However, this density matrix can be decomposed in an infinite number of ways, i.e.\ it can describe an infinite number of ensembles.  We have already imposed some conditions, that the ensemble consist of two states whose probabilities of occurrence we know.  That, however, is not sufficient.  There are still many choices of pairs of states with the assigned probabilities that will yield the same density matrix.  Ideally we would like conditions that specify the ensemble uniquely.  If this is not the case, we can still sometimes do something.  For example, if we have imposed conditions so that there are only two possible decompositions (this situation will subsequently occur),
\begin{eqnarray}
\rho & = & \eta_{0} |\psi_{0}\rangle\langle \psi_{0}| + \eta_{1} |\psi_{1}\rangle\langle\psi_{1}|
\nonumber \\
 & = &  \eta_{0} |\phi_{0}\rangle\langle \phi_{0}| + \eta_{1} |\phi_{1}\rangle\langle\phi_{1}| ,
\end{eqnarray}
and we don't know which of the ensembles $\{ |\psi_{j}\rangle \, |\, j=0,1 \}$ or $\{ |\phi_{j}\rangle \, |\, j=0,1 \}$ we have, we cannot distinguish them by making measurements, because they have the same density matrix.  However, if $|\psi_{0}\rangle$ is is close to $|\phi_{0}\rangle$ and $|\psi_{1}\rangle$ is close to $|\phi_{1}\rangle$, then we may be able to find a POVM that does a reasonable job for both cases.

We will consider two cases, the the case in which the Bloch vectors of $|\psi_{0}\rangle$ and $|\psi_{1}\rangle$ lie in the $x-z$ plane of the Bloch sphere, and the case in which their Bloch vectors have the same $z$ component.  The second case is a generalization of the first.  In the first case this implies that we know the vectors $|0\rangle$ and $|1\rangle$, and that $|\psi_{0}\rangle$ and $|\psi_{1}\rangle$ are linear combinations of them with real coefficients.  In the second case, we also know $|0\rangle$ and $|1\rangle$, and the states all yield the same expectation of $\sigma_{z}$.  We will first consider the case in which the two states are equally probable, and then move on to the case in which they are not.

\section{States in the $x-z$ plane}
\subsection{Equally probable states}

The density matrix describing our sequence, or ensemble, is now 
\begin{equation}
\label{density}
\rho = \frac{1}{2}( |\psi_{0}\rangle\langle \psi_{0}| +  |\psi_{1}\rangle\langle \psi_{1}| ) .
\end{equation} 
The Bloch vector for $\rho$, $\mathbf{n}$, is equal to $(\mathbf{n}_{0}+\mathbf{n}_{1})/2$, where $\mathbf{n}_{0}$ and $\mathbf{n}_{1}$ are the Bloch vectors for $|\psi_{0}\rangle\langle\psi_{0}|$ and $|\psi_{1}\rangle\langle\psi_{1}|$, respectively.  We assume we have determined $\mathbf{n}$  by performing measurements on the ensemble.  All three Bloch vectors lie in the $x-z$ plane, and $\mathbf{n}_{0}$ and $\mathbf{n}_{1}$ are of length one, because they correspond to pure states.

It is relatively straightforward to see that in this situation each density matrix corresponds to a unique pair of states.  We can express $|\psi_{0}\rangle$ and $|\psi_{1}\rangle$ as 
\begin{eqnarray}
|\psi_{0}\rangle & = & \cos \left( \frac{\alpha + \beta}{2}\right) |0\rangle + \sin \left( \frac{\alpha + \beta}{2} \right) |1\rangle \nonumber \\
|\psi_{1}\rangle & = & \cos \left( \frac{\alpha - \beta}{2} \right) |0\rangle + \sin \left( \frac{\alpha - \beta}{2} \right) |1\rangle  ,
\end{eqnarray}
where $0\leq \beta \leq \pi /2$ and $0\leq \alpha \leq 2\pi$.  The corresponding Bloch vectors are $\mathbf{n}_{0}=(\sin (\alpha + \beta ), \cos (\alpha + \beta ))$ and $\mathbf{n}_{1}=(\sin (\alpha - \beta ), \cos (\alpha - \beta ))$, where we have listed the $x$ oomponent first and the $z$ component second. For $\mathbf{n}$ we have $\mathbf{n} = (\sin\alpha  , \cos\alpha ) \cos\beta$.  We see that $\mathbf{n}$ determines both $\beta$ and $\alpha$; its magnitude is $\cos\beta$ and its orientation is given by $\alpha$.  Once these angles are determined, so are $|\psi_{0}\rangle$ and $|\psi_{1}\rangle$.

We now want a POVM that will distinguish $|\psi_{0}\rangle$ and $|\psi_{1}\rangle$.  From the theory of state discrimination, we know that the optimal POVM for minimum error discrimination of two states is a projective measurement \cite{helstrom}.  That means we can take the POVM elements to be $\Pi_{0} = |v_{0}\rangle\langle v_{0}|$ and $\Pi_{1} = |v_{1}\rangle\langle v_{1}|$, where
\begin{eqnarray}
\label{povm1}
|v_{0}\rangle & = & \cos\phi |0\rangle + \sin\phi |1\rangle \nonumber \\
|v_{1}\rangle & = & \sin\phi |0\rangle - \cos\phi |1\rangle .
\end{eqnarray}
We want $\Pi_{0}$ to correspond to detecting $|\psi_{0}\rangle$ and $\Pi_{1}$ to correspond to detecting $|\psi_{1}\rangle$.  The optimal POVM results when $\phi = (\alpha /2)+\pi /4$.

The question now is how can we extract $\alpha$ by measuring states in the sequence.  One method is simply to perform state tomography measurements in order to determine $\rho$, and thereby $\alpha$. Here we would like to explore another method.  Let us find $p_{0}$, the probability that we obtain the result $0$ with the above POVM and $p_{1}$, the probability that we obtain $1$.  We have
\begin{eqnarray}
p_{0} & = &  \frac{1}{2} ( |\langle v_{0}|\psi_{0}\rangle |^{2} + |\langle v_{0}|\psi_{1}\rangle |^{2} ) \nonumber \\
& = & \frac{1}{2} \left[ \cos^{2}\left( \frac{\alpha + \beta}{2} -\phi \right) + \cos^{2}\left( \frac{\alpha - \beta}{2} -\phi \right) \right] , \nonumber \\
\end{eqnarray}
and
\begin{eqnarray}
p_{1} & = & \frac{1}{2} ( |\langle v_{1}|\psi_{0}\rangle |^{2} + |\langle v_{1}|\psi_{1}\rangle |^{2} ) \nonumber \\
 & = & \frac{1}{2} \left[ \sin^{2}\left( \frac{\alpha + \beta}{2} -\phi \right) + \sin^{2}\left( \frac{\alpha - \beta}{2} -\phi \right) \right]  . \nonumber \\
\end{eqnarray}
The difference between the two quantities is
\begin{equation}
p_{0}-p_{1} = \cos (\alpha - 2\phi ) \cos (\beta ) ,
\end{equation}
and this is zero when $\phi = (\alpha /2)+\pi /4$, the value of $\phi$ that yields the optimal POVM.  Therefore, we can tune $\phi$ by finding a value where the detectors are equally likely to fire.

This can be done as follows.  To find the proper value of $\phi$ for given values of $\alpha$ and $\theta$, let $\Delta (\phi ) = p_{0}-p_{1}$  Choose two values of $\phi$, $\phi_{0}$ and $\phi_{0} + \pi /4$, where $\phi_{0}$ is arbitrary.  We then have that
\begin{equation}
\frac{\Delta (\phi_{0} + \pi /4)}{\Delta (\phi_{0})} = \tan (\alpha - 2\phi_{0}) .
\end{equation}
Note that $\Delta (\phi_{0})$ and $\Delta (\phi_{0} + \pi /4)$ can be measured.  From the above equation we can solve for $\alpha$, since we know $\phi_{0}$, and with this knowledge we can then set $\phi$ to the proper value, $\alpha + \pi /4$, for the optimal POVM. Note that if $\beta$ is too close to $\pi /2$, the dependence of $\Delta (\phi )$ on $\phi$ will be weak, which will make the determination of $\alpha$ difficult.

Before proceeding, let us discuss further the condition we imposed that the Bloch vectors of the two states lie in the $x-z$ plane.  We had the relation $\mathbf{n} = (\mathbf{n}_{0}+\mathbf{n}_{1})/2$, relating the Bloch vectors of the sequence to the Bloch vectors of the states that make it up.  If we now rotate $\mathbf{n}_{0}$ and $\mathbf{n}_{1}$ about $\mathbf{n}$ by some angle, to get $\mathbf{n}_{0}^{\prime}$ and $\mathbf{n}_{1}^{\prime}$, we will have that $\mathbf{n} = (\mathbf{n}_{0}^{\prime} +\mathbf{n}_{1}^{\prime} )/2$.  Therefore, there is a circle of states on the surface of the Bloch sphere into which our density matrix, $\rho$, can be decomposed.  We can make the decomposition unique by specifying the plane in which $\mathbf{n}_{0}$ and $\mathbf{n}_{1}$ must lie.

\subsection{States with different  probabilities}

We will now return to requiring the Bloch vectors to be in the $x-z$ plane, but allow the probabilities, $\eta_{0}$ and $\eta_{1}$, to be different.  As we shall see, this immediately introduces a complication into the problem.  The decomposition of a given sequence density matrix is no longer unique; there are, in fact, two possible decompositions for each density matrix.

If $\mathbf{n}$ is the Bloch vector for the ensemble density matrix, and $\mathbf{n}_{0}$ and $\mathbf{n}_{1}$ are the Bloch vectors of the pure states in the sequence, we have
\begin{equation}
\label{n}
\mathbf{n} = \eta_{0}\mathbf{n}_{0} + \eta_{1} \mathbf{n}_{1} .
\end{equation}
Taking the inner product of each side of this equation with itself and denoting the angle between $\mathbf{n}_{0}$ and $\mathbf{n}_{1}$ by $\theta$, we find
\begin{equation}
\cos\theta = \frac{|\mathbf{n}|^{2} - \eta_{0}^{2} - \eta_{1}^{2} }{2\eta_{0}\eta_{1}} .
\end{equation}
If we now let $\alpha$ be the angle between $\mathbf{n}_{0}$ and the $x$ axis, we then have two cases. Either the angle $\mathbf{n}_{1}$ makes with the $x$ axis is $\alpha -\theta$, which we shall call case $A$, or the angle it makes with the $x$ axis is $\alpha + \theta$, which we shall call case $B$.  Therefore, we have for case $A$
\begin{eqnarray}
n_{1x} & = & \cos (\alpha - \theta ) = n_{0x} \cos\theta + n_{0z} \sin\theta \nonumber \\
n_{1z} & = & \sin (\alpha - \theta ) = n_{0z} \cos\theta - n_{0x} \sin\theta ,
\end{eqnarray}
and for case $B$,
\begin{eqnarray}
n_{1x} & = & \cos (\alpha + \theta ) = n_{0x} \cos\theta - n_{0z} \sin\theta \nonumber \\
n_{1z} & = & \sin (\alpha + \theta ) = n_{0z} \cos\theta + n_{0x} \sin\theta ,
\end{eqnarray}
Inserting these expressions into Eq.\ (\ref{n}) and solving for $\mathbf{n}_{0}$ and $\mathbf{n}_{1}$ in terms of $\mathbf{n}$, we have for case $A$ that
\begin{equation}
\label{A0}
\left( \begin{array}{c} n_{0x} \\ n_{0z} \end{array} \right) =  \frac{1}{|\mathbf{n}|^{2}} \left( \begin{array}{cc} \eta_{0}+\eta_{1}\cos\theta & -\eta_{1}\sin\theta \\ \eta_{1}\sin\theta & \eta_{0}+\eta_{1}\cos\theta \end{array} \right)  \left( \begin{array}{c} n_{x} \\n_{z} \end{array} \right) ,
\end{equation} 
and
\begin{equation}
\left( \begin{array}{c} n_{1x} \\ n_{1z} \end{array} \right) =  \frac{1}{|\mathbf{n}|^{2}} \left( \begin{array}{cc} \eta_{1}+\eta_{0}\cos\theta & \eta_{0}\sin\theta \\ -\eta_{0}\sin\theta & \eta_{1}+\eta_{0}\cos\theta \end{array} \right)  \left( \begin{array}{c} n_{x} \\n_{z} \end{array} \right) .
\end{equation} 
For case $B$ we have
\begin{equation}
\left( \begin{array}{c} n_{0x} \\ n_{0z} \end{array} \right) =  \frac{1}{|\mathbf{n}|^{2}} \left( \begin{array}{cc} \eta_{0}+\eta_{1}\cos\theta & \eta_{1}\sin\theta \\ -\eta_{1}\sin\theta & \eta_{0}+\eta_{1}\cos\theta \end{array} \right)  \left( \begin{array}{c} n_{x} \\n_{z} \end{array} \right) ,
\end{equation} 
and 
\begin{equation}
\label{B1}
\left( \begin{array}{c} n_{1x} \\ n_{1z} \end{array} \right) =  \frac{1}{|\mathbf{n}|^{2}} \left( \begin{array}{cc} \eta_{1}+\eta_{0}\cos\theta & -\eta_{0}\sin\theta \\ \eta_{0}\sin\theta & \eta_{1}+\eta_{0}\cos\theta \end{array} \right)  \left( \begin{array}{c} n_{x} \\n_{z} \end{array} \right) .
\end{equation} 

From these expressions, we note that $\mathbf{n}_{0}^{A} - \mathbf{n}_{1}^{B}$ goes to zero as $\eta_{0}$ and $\eta_{1}$ approach $1/2$, as does  $\mathbf{n}_{1}^{A} - \mathbf{n}_{0}^{B}$, where we have indicated to which case the vectors belong by superscripts.  If $|\eta_{0}- \eta_{1}|$ is small, then we can still hope to find a POVM that will be able to distinguish the states in the sequence.  This POVM will have two elements, $\Pi_{0}$ and $\Pi_{1} = I - \Pi_{0}$,  $\Pi_{0}$ corresponds to detecting either $\rho_{0}^{A}$ or $\rho_{1}^{B}$, and $\Pi_{1}$ corresponds to detecting $\rho_{1}^{A}$ or $\rho_{0}^{B}$, where $\rho_{0}^{A}$ is the state with the Bloch vector $\mathbf{n}_{0}^{A}$, and similarly for the other states.  Note that all four of these states, that is $\rho_{0}^{A}$, $\rho_{1}^{A}$, $\rho_{0}^{B}$, and $\rho_{1}^{B}$, are pure state density matices.  Let us assume that the cases $A$ and $B$ are equally probable.  We then have that the probability of $\rho_{0}^{A}$ occurring is $(1/2)\eta_{0}$, and the probability of successfully detecting it if it does occur is ${\rm Tr}(\Pi_{0}\rho_{0}^{A})$, with similar expressions for the other states.  Therefore, the probability of successfully identifying a states, $P_{s}$, is
\begin{eqnarray}
P_{s} & = & \frac{1}{2} [ \eta_{0} {\rm Tr}(\Pi_{0}\rho_{0}^{A}) + \eta_{1} {\rm Tr}(\Pi_{1}\rho_{1}^{A}) \nonumber \\
& & + \eta_{1}  {\rm Tr}(\Pi_{0}\rho_{1}^{B}) + \eta_{0} {\rm Tr}(\Pi_{1}\rho_{0}^{B})  ],  \nonumber \\
& = &  \frac{1}{2} [  {\rm Tr}(\Pi_{0}\rho_{0}) +  {\rm Tr}(\Pi_{1}\rho_{1}) ] ,
\end{eqnarray}
where
\begin{eqnarray}
\rho_{0} & = & \eta_{0} \rho_{0}^{A} + \eta_{1}\rho_{1}^{B} \nonumber \\
\rho_{1} & = & \eta_{1} \rho_{1}^{A} + \eta_{0} \rho_{0}^{B} .
\end{eqnarray}
The Bloch vectors corresponding to $\rho_{0}$ and $\rho_{1}$ are $\mathbf{m}_{0}$ and $\mathbf{m}_{1}$, respectively, where
\begin{eqnarray}
m_{0x} & = & \frac{1}{|\mathbf{n}|^{2}} \left[ n_{x} |\mathbf{n}|^{2} -2\eta_{0}\eta_{1} n_{z} \sin\theta  \right]\nonumber \\ 
m_{0z} & = & \frac{1}{|\mathbf{n}|^{2}} \left[2\eta_{0}\eta_{1} n_{x}\sin\theta + n_{z} |\mathbf{n}|^{2} \right] , 
\end{eqnarray}
and
\begin{eqnarray}
m_{1x} & = & \frac{1}{|\mathbf{n}|^{2}} \left[ n_{x} |\mathbf{n}|^{2} + 2\eta_{0}\eta_{1} n_{z} \sin\theta  \right]\nonumber \\ 
m_{1z} & = & \frac{1}{|\mathbf{n}|^{2}} \left[ -2\eta_{0}\eta_{1} n_{x}\sin\theta + n_{z} |\mathbf{n}|^{2} \right] . 
\end{eqnarray}

Optimizing $P_{s}$ is just the problem of optimally discriminating the two equiprobable density matrices, $\rho_{0}$ and $\rho_{1}$, with minimum error, and the solution to this problem is well known \cite{helstrom}.  As noted before, the optimal measurement is a projective measurement, so $\Pi_{0}$ and $\Pi_{1}$ are projections onto orthogonal pure states.  Let $\mathbf{s}$ be the Bloch vector corresponding to the projection operator $\Pi_{0}$ and $-\mathbf{s}$ correspond to the orthogonal projector $\Pi_{1}$.  For $\mathbf{s}$ we have that $|\mathbf{s}|=1$, since $\Pi_{0}$ is a one-dimensional projection, and it must lie in the $x-z$ planed (see the discussion in the appendix).  One way of determining $\mathbf{s}$ is by the condition that if the purity of the two density matrices to be discriminated is the same, i.e.\ if ${\rm Tr} (\rho_{0}^{2}) = {\rm Tr}(\rho_{1}^{2})$, which is true in this case, then for the optimal measurement the number of counts in detector $0$ will be the same as the number of counts in detector $1$.  See the appendix for a discussion of this point.  The probability that detector $0$ will click is
\begin{eqnarray}
p_{0} & = & \frac{1}{2} [{\rm Tr}(\Pi_{0}\rho_{0}) + {\rm Tr}(\Pi_{0}\rho_{1}) ]  \nonumber \\
& = & {\rm Tr }(\Pi_{0}\rho ) ,
\end{eqnarray}
where $\rho$ is the density matrix of the ensemble.  Similarly, $p_{1}$, the probability of detector $1$ firing is ${\rm Tr}(\Pi_{1}\rho )$.  In terms of Bloch vectors, we have
\begin{eqnarray}
\label{detect-prob}
p_{0} & = & \frac{1}{2} (1 +  \mathbf{s}\cdot \mathbf{n}) \nonumber \\
p_{1} & = & \frac{1}{2} (1 - \mathbf{s} \cdot \mathbf{n}) .
\end{eqnarray}
These will be equal when $\mathbf{s} \cdot \mathbf{n} = 0$.  Therefore, the optimal measurement can be found by measuring $\sigma_{x}$ and $\sigma_{z}$, since ${\rm Tr}(\sigma_{x}\rho )=n_{x}$ and ${\rm Tr}(\sigma_{z}\rho ) = n_{z}$, and finding the orthogonal projection operators whose Bloch vectors are orthogonal to $\mathbf{n}$.

In more detail, defining
\begin{equation}
\mathbf{n}^{\perp} = \frac{1}{|\mathbf{n}|} \left( \begin{array}{c} -n_{z} \\ n_{x} \end{array} \right) ,
\end{equation}
we have that
\begin{eqnarray}
\mathbf{m}_{0} & = & \mathbf{n} + \frac{2\eta_{0}\eta_{1}\sin\theta}{|\mathbf{n}|} \mathbf{n}^{\perp} 
\nonumber \\
\mathbf{m}_{1} & = & \mathbf{n} - \frac{2\eta_{0}\eta_{1}\sin\theta}{|\mathbf{n}|} \mathbf{n}^{\perp} .
\end{eqnarray}
From this we see that we should choose $\mathbf{s} = \mathbf{n}^{\perp}$, and that the success probability will be 
\begin{equation}
P_{s}= \frac{1}{2} + \frac{\eta_{0}\eta_{1}\sin\theta}{|\mathbf{n}|} .
\end{equation}
For any value of $\theta$, $P_{s}$ is a maximum when $\eta_{0} = \eta_{1} = 1/2$.

\section{States in a plane of constant $z$}
So far we considered states in the $x-z$ plane and found a POVM that would allow us to discriminate between the states in our ensemble.  This implies that a similar procedure would work for states lying in any plane that passed through the origin of the Bloch sphere.  We could simply redefine our coordinate system so that that plane became the $x-z$ plane.  

We can also treat a different case.  Suppose we know that the Bloch vectors of the two states making up the ensemble, $\mathbf{n}_{0}$ and $\mathbf{n}_{1}$, have the same $z$ component, $n_{z}$.  Then, their linear combination $\mathbf{n}=\eta_{0}\mathbf{n}_{0} + \eta_{1}\mathbf{n}_{1}$ also has the same $z$ component, since $\eta_{0}+\eta_{1}=1$.  We can express the Bloch vectors as $\mathbf{n}= \mathbf{r} + n_{z}\mathbf{\hat{z}}$ and $\mathbf{n}_{j}= \mathbf{r}_{j} + n_{z}\mathbf{\hat{z}}$, for $j=0,1$, where the vectors $\mathbf{r}$, $\mathbf{r}_{0}$ and $\mathbf{r}_{1}$ lie in the $x-y$ plane.  We assume that we have determined $\mathbf{r}$ by performing measurements on the ensemble.  We have that
\begin{equation}
\mathbf{r} = \eta_{0}\mathbf{r}_{0} + \eta_{1}\mathbf{r}_{1} ,
\end{equation}
and that $|\mathbf{r}_{0}|=|\mathbf{r}_{1}| = (1-n_{z}^{2})^{1/2}$.  In this case we have for $\theta$, the angle between $\mathbf{r}_{0}$ and $\mathbf{r}_{1}$, 
\begin{equation}
\cos\theta = \frac{1}{\eta_{0}\eta_{1}} \left( \frac{|\mathbf{r}|^{2}}{1-n_{z}^{2}} -\eta_{0}^{2} - \eta_{1}^{2} \right) .
\end{equation}

As before we have two cases.  We let $\alpha$ be the angle between $\mathbf{r}_{0}$ and the $x$ axis.  Either the angle $\mathbf{r}_{1}$ makes with the $x$ axis is $\alpha - \theta$, case A, or $\alpha + \theta$, case B. The relations between $\mathbf{r}_{0}$, $\mathbf{r}_{1}$ and $\mathbf{r}$ are given by Eqs.\ (\ref{A0} - \ref{B1}), with $1/|\mathbf{n}|^{2}$ replaced by $(1-n_{z}^{2})/|\mathbf{r}|^{2}$.  

Similar to before, as $\eta_{0}$ and $\eta_{1}$ approach $1/2$, $\mathbf{r}^{A}_{0}-\mathbf{r}_{1}^{B}$ and $\mathbf{r}_{1}^{A} - \mathbf{r}_{0}^{B}$ approach zero.  Following the previous discussion, that means we want a two element POVM that will discriminate density matrices with Bloch vectors 
\begin{eqnarray}
\mathbf{m}_{0} & = & \eta_{0} \mathbf{r}_{0}^{A} + \eta_{1} \mathbf{r}_{1}^{B} + n_{z}\mathbf{\hat{z}} \nonumber \\
\mathbf{m}_{1} & = & \eta_{1} \mathbf{r}_{1}^{A} + \eta_{0} \mathbf{r}_{0}^{B} + n_{z}\mathbf{\hat{z}} ,
\end{eqnarray}
The POVM elements will be projection operators, and we can take the Bloch vector corresponding to $\Pi_{0}$, which corresponds to detecting the density matrix with the Bloch vector $\mathbf{m}_{0}$, to be $\mathbf{s}$, and $\Pi_{1}$, corresponding to detecting the density matrix with Bloch vector $\mathbf{m}_{1}$, to be $-\mathbf{s}$.  The vector $\mathbf{s}$ is a unit vector and lies in the $x-y$ plane (see discussion in the appendix).  The probabilities that $\Pi_{0}$ clicks, $p_{0}$, and that $\Pi_{1}$ clicks, $p_{1}$, are
\begin{eqnarray}
p_{0} & = & \frac{1}{2}(1 + \mathbf{s}\cdot\mathbf{r} ) \nonumber \\
p_{0} & = & \frac{1}{2}(1 - \mathbf{s}\cdot\mathbf{r} )  .
\end{eqnarray}
For these to be equal, we need $ \mathbf{s}\cdot\mathbf{r}  = 0$, which, since both vectors are in the $x-y$ plane, determines $\mathbf{s}$ up to a sign.

This result can be generalized by just rotating the configuration.  Choose a radius vector in the Bloch sphere and construct a plane perpendicular to it.  This plane intersects the Bloch sphere in a circle and its interior.  States with Bloch vectors ending on the circle and its interior can be treated by the procedure above.  Just call the radius vector the $z$ axis and proceed.  Note that the earlier case we considered, states in the $x-z$, plane, is a special case of the case presented in this section.

\section{Conclusion}
We have shown that for some ensembles of qubits consisting of two different pure states, $|\psi_{0}\rangle$ and $|\psi_{1}\rangle$, it is possible to construct a POVM that will, imperfectly, discriminate between the two states.  The POVM can then be used to separate the remainder of the data into two groups, one group consisting primarily of qubits that were in the state $|\psi_{0}\rangle$, and the other consisting of states that were in the state $|\psi_{1}\rangle$.

\section*{Appendix}
Suppose $\rho_{0}$ and $\rho_{1}$ are two qubit density matrices, and each occurs with a probability of $1/2$.  That is, we are given a qubit, and it is equally likely to be in the state $\rho_{0}$ or $\rho_{1}$.  Our task is then to determine the state of the qubit using a measurement that minimizes the probability of making a mistake.  The measurement is a projective one, and it is found by diagonalizing the operator $\rho_{0} - \rho_{1}$ \cite{helstrom}.  We can express this operator as
\begin{equation}
\rho_{0} - \rho_{1} = \lambda (P_{0} - P_{1} )  ,
\end{equation}
where shall assume that $\lambda > 0$, $P_{0}$ and $P_{1}$ are one-dimensional projections, and $P_{0} + P_{1} =I$.  
Note that the eigenvalues of $\rho_{0} - \rho_{1}$ are $\pm \lambda$, because the trace of the operator is zero.  $P_{0}$ is the detection operator for $\rho_{0}$ and $P_{1}$ is the detection operator for $\rho_{1}$.  We can solve the above equation for $P_{0}$ and $P_{1}$ in terms of $\rho_{0}$ and $\rho_{1}$
\begin{eqnarray}
P_{0} & = & \frac{1}{2} \left[ I + \frac{1}{\lambda} (\rho_{0} - \rho_{1}) \right] \nonumber \\
P_{1} & = & \frac{1}{2} \left[ I - \frac{1}{\lambda} (\rho_{0} - \rho_{1}) \right]  .
\end{eqnarray}
The probability of triggering detector $0$ is $(1/2){\rm Tr}(P_{0} (\rho_{0} + \rho_{1}))$ and the probability of triggering detector $1$ is ${\rm Tr}(P_{1} (\rho_{0} + \rho_{1}))$.  These will be equal if 
\begin{equation}
{\rm Tr}((\rho_{0}-\rho_{1})(\rho_{0} + \rho_{1})) = {\rm Tr}(\rho_{0}^{2}) - {\rm Tr}(\rho_{1}^{2}) =0.
\end{equation}
If $\mathbf{m}_{0}$ and $\mathbf{m}_{1}$ are the Bloch vectors for $\rho_{0}$ and $\rho_{1}$, respectively, the above condition is equivalent to $|\mathbf{m}_{0}| = |\mathbf{m}_{1}|$, which is satisfied in the cases considered in the paper.

So far we have shown that the optimal measurement satisfies the condition that the probability to trigger both detectors is the same.  We would like to show the converse.  We first note that in the first case, the Bloch vector corresponding to $\rho_{0}-\rho_{1}$ lies in the $x-z$ plane.  It is relatively straightforward to show that this implies that the Bloch vectors for $P_{0}$ and $P_{1}$ also lie in the $x-z$ plane.  Similarly, in the second case considered, the Bloch vector corresponding to $\rho_{0}-\rho_{1}$ lies in the $x-y$ plane, and this implies that the Bloch vectors for $P_{0}$ and $P_{1}$ also lie in the $x-y$ plane.  For both cases, we found that the condition for equal detection probabilities was $\mathbf{s} \cdot \mathbf{n} = 0$, where both vectors are in the same plane.  In a two dimensional space this condition determines $\mathbf{s}$, the Bloch vector corresponding to the measurement operator, up to a sign..  Therefore, in the cases we considered, we can use this condition to determine the optimal measurement.


\begin{thebibliography}{99}
\bibitem{bisio} A.~Bisio, G.~Chiribella, G.~M.~D'Ariano, S.~Facchini, and P.~Perinotti, Phys.\ Rev.\ A {\bf 81}, 032324 (2010).
\bibitem{sedlak} A.~Bisio, G.~M.~D'Ariano, S.~Facchini, P.~Perinotti, and M.~Sedlak, Phys.\ Lett.\ A {\bf 375}, 3425 (2011).
\bibitem{guta} M.~Guta and W.~Kotlowski, New J.\ Phys.\ {\bf 12}, 123032 (2010).
\bibitem{sentis} G.~Sentis, J.~Calsamglia, R.~Munoz-Tapia, and E.~Bagan, Sci.\ Rep.\ {\bf 2}, 708 (2012).
\bibitem{fanizza} M.~Fanizza, A.~Mari, and V.~Giovannetti, IEEE Trans.\ Inf.\  Theory {\bf 65}, 5931 (2019).
\bibitem{Brassard} E.~Aimeur, G.~Brassard, and S.~Gambs, Machine Learning {\bf 90}, 261 (2013)
\bibitem{Lloyd2} S.~Lloyd, M.~Mohseni, and P.~Rebentrost, arXiv:1307.0411.
\bibitem{Svore} N.~Wiebe, A.~Kapoor, and K.~Svore, Quantum Information and Computation {\bf 15}, 0318 (2015).
\bibitem{kerenidas} I.\ Kerenidas, J.\ Landman, A.\ Luongo, and A.\ Prakash, Advances in Neural Information Processing Systems {\bf 32}, NICS (2019) and arXiv:1812.03584 .
\bibitem{schuld} M.~Schuld, I.~Sinayskiy, and F.~Petruccione, Contemporary Physics {\bf 56}, 172 (2015).
\bibitem{bagan} G.\ Sentis, A.\ Monras, R.\ Munoz-Tapia, J.\ Calsamiglia, and E.\ Bagan, Phys.\ Rev.\ X {\bf 9}, 041029 (2019). 
\bibitem{helstrom} C.~W.~Helstrom, \emph{Quantum Detection and Estimation Theory} (Academic, New York, 1976).
\bibitem{bergou} For a review of state discrimination see Discrimination of Quantum States by J.~A.~Bergou, U.~Herzog, and M.~Hillery in \emph{Quantum State Estimation}, edited by M.~G.~A.~Paris and J.~\v{R}eha\v{c}ek (Springer Verlag, Berlin, 2004).
\end{thebibliography}
\end{document}